# Model Rocket Project for Aerospace Engineering Course: Trajectory Simulation and Propellant Analysis

Thomas A. Campbell Jr., Masataka Okutsu

*Department of Mechanical Engineering*
*The Catholic University of America, Washington D.C., USA*

Model rockets have been employed in student projects, but very few papers in aerospace education offer concise summaries of activities at university-course levels. This paper aims to address this gap in the literature. The rockets used by our students reached some 500 m (~1,640 feet) in altitude, deployed a parachute, and spent 2–3 minutes descending to the ground. We present a series of analyses and experiments that students performed in order to predict the flight time, the maximum altitude, and the landing location of these rockets. They wrote computer programs to numerically integrate equations of motion, and experimentally measured input parameters (e.g., the thrust profile and drag coefficients). Once launched, these rockets could not be controlled; targeting the landing location would thus mean tilting the launch rail to a required angle. The largest source of error in landing location came from the difficulty in modeling wind velocities. Also discussed in this paper are the infrared spectroscopy and the extraction experiment as novel additions to model rocket projects.

**Nomenclature**

| | |
|---|---|
| $A$ | area, m$^2$ |
| $a$ | Hellmann exponent |
| $C_D$ | drag coefficient |
| $D$ | drag, N |
| $d$ | distance, m |
| $g$ | standard free fall acceleration, i.e., 9.80665 m/s$^2$ |
| $h$ | enthalpy, kJ/kg |
| $I_{sp}$ | specific impulse, s |
| $I_{total}$ | total impulse, N·s |
| $L_0$ | temperature lapse rate with respect to altitude, K/m |
| $m$ | mass, kg |
| $M$ | molar mass, kg/mol |
| $R$ | universal gas constant, i.e., 8.3144598 J K$^{-1}$ mol$^{-1}$ |
| $T$ | thrust, N |
| $t$ | time, s |
| $T_0$ | temperature of air at the ground level, K |
| $V$ | velocity, m/s |
| $y$ | altitude measured from ground, m |
| $Y$ | altitude measured from sea level, m |
| $\alpha$ | angle of attack, radians |
| $\varepsilon_{kinetic}$ | specific kinetic energy, kJ/kg |
| $\eta$ | thermochemical efficiency |
| $\theta$ | pitch angle, radians |
| $\gamma$ | flight path angle, radians |
| $\rho$ | density of air, kg/m$^3$ |
| $\rho_0$ | density of air at the ground level, kg/m$^3$ |
| $\varphi$ | horizontal angle between the rocket and the other observer, radians |
| $\psi$ | vertical angle of elevation between the ground and the rocket, radians |

## 1. Introduction

Projects on model rockets offer opportunities for analyses and experiments [1]. In the literature of education, however, papers on this topic either focus on one specific technical area (while ignoring the multi-disciplinary nature of the endeavor) or merely introduce the projects as educational-outreach activities (while ignoring learning objectives for engineering students). Very few papers offer a summary of technical activities suitable at university levels.

The lack of literature on model rocket projects can be exemplified by the database *Aerospace Research Central* (ARC), which archives all papers published by the American Institute of Aeronautics and Astronautics (AIAA)—the world's largest professional organization for aerospace engineers. Entering such keywords as *model rockets* and *rocket projects* on the ARC database would result in only four publications; two of which are our projects (i.e., Campbell et al. [1] and Brewer et al. [2]) that provide foundation for this work.

Going beyond the ARC database, we made use of *Google Scholar* (Google) and *Academic Search Complete* (EBSCOhost). Using a similar set of keywords, we conducted multiple search trials and reviewed the first hundred results on each search.

Many papers identified in this process would address only one technical aspect of model rocket projects. For example, Haw [3], Penn and Slaton [4], and Dooling [5] discuss an apparatus or procedure to measure rocket motors' thrust; and Caughey [6] and Nelson and Wilson [7] discuss equations of motion (EOM) for the rocket trajectory.

Other papers may concern either different categories of rockets or consider model rockets merely as a tool to introduce engineering concepts. For example, Stancato, Mangili, and Destro [8] discuss experimental rocket projects for first-year university students (who only had high-school level course work). Papers that concern students in high schools or at lower grade levels include the following: Tomita, Watanabe, and Nebylov [9] discussing a water rocket project; Villarreal et al. [10] discussing model rocket projects as an outreach activity (to K–12 students); Horst [11] discussing trajectory and flight data; Keith, Martin, and Veltkamp [12] discussing a model rocket project; Elger, Beyerlein, and Budwig [13] discussing model rockets as a design, build, and test (DBT) project.

Among the papers uncovered in our search, Jayaram et al. [14] was the only one that catalogued analyses and experiments suitable for university students. While the target students of the project described by Jayaram et al. were in their first or second years of the undergraduate curriculum, our target students in this paper are in their third or fourth years.

The rockets employed in our project (Fig. 1) would reach an altitude up to 500 m (~1,640 feet), which is much higher than 30 m (100 ft) reached by rockets in the project as described by Jayaram et al. Our rockets are still classified as Class 1 *Model Rockets* (with the rocket weighing less than 16 oz. and with the propellant weighing less than 5 oz.), thus eliminating the need for a special FAA wavier.

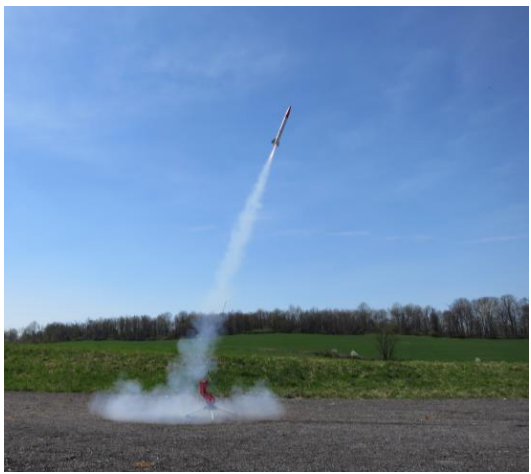

**Fig. 1.** The model rocket *Ascender* is launched into the wind.

Of course, some sounding rockets could reach much higher altitudes (e.g., > 50 km). But such projects would require special launch site, trained supervisors, and funding levels far greater than those allocated for typical university courses. In contrast, our aim is to catalogue activities that could be readily adopted at most engineering schools.

Page 3 of 22

Our students were asked to predict the maximum altitude, the landing location, and the time of flight. To make these predictions they wrote computer programs to numerically integrate equations of motion, and conducted experiments to measure the input parameters (e.g., thrust of rocket motor, drag coefficient of parachute). They also estimated the efficiency of the rocket motor through thrust testing and chemistry experiments. A multi-disciplinary approach to the project allowed the students to synthesize knowledge they gained throughout the undergraduate curriculum, including from such courses as fluid mechanics, dynamics, thermodynamics, chemistry, differential equations, and computer programming.

Due to concerns for safety, we elected to use off-the-shelf rockets and motors, rather than fabricating them from scratch. The hardware was purchased from Estes, one of the main manufacturers of model rockets in the United States. To satisfy the instructor's requirement—that the rockets would not require special license, but could reach an altitude of at least 300 m (approximately 1000 feet)—our student teams selected the *Ascender*, the *Magician*, and the *Power Patrol* rockets; to power them, they selected the F15-6, the E9-6, and the C6-5 motors, respectively. Unless stated otherwise, examples used in this paper correspond to the *Magician* rocket powered by the E9-6 motor.

## 2. Simulation of flight trajectories

Students were tasked to predict the time of flight, the maximum altitude, and the landing location. Instead of the landing location, they could instead predict the launch angle that would make the rockets return to the launch pad.

*2.1. Phases of model rocket trajectories*

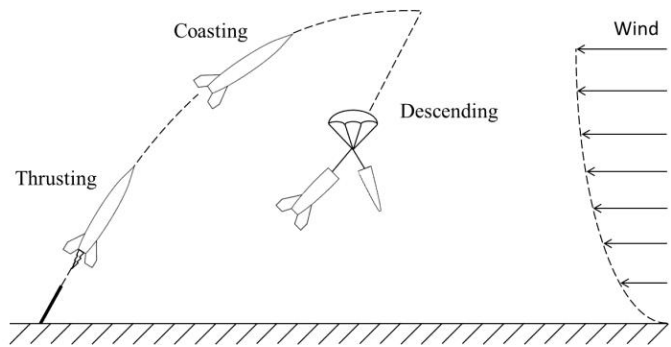

**Fig. 2.** Model rocket flight phases.

The flight trajectory of our model rocket consists of three phases (Fig. 2). The thrusting phase corresponds to the first 3 s of the flight when the rocket motor burns, the coasting phase corresponds to the next 5 s of the flight when the rocket continues to fly without thrusting, and the descending phase corresponds to the remaining 1–3 minutes of the flight when the rocket slowly descends via parachute. The goal of the project is to simulate these three phases.

## 2.2. Forces on rocket body

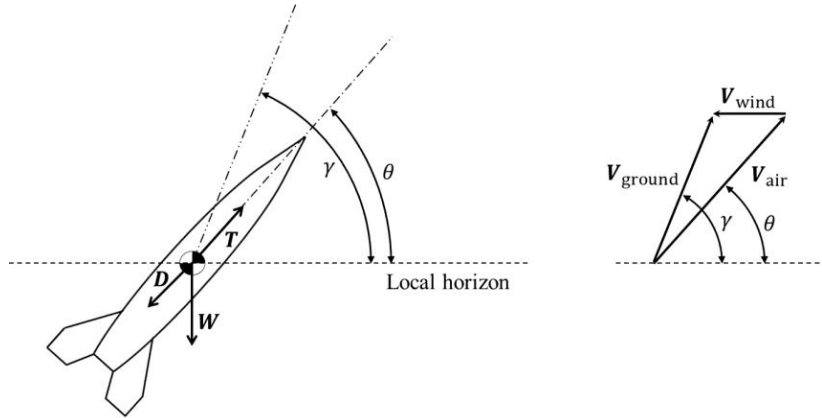

**Fig. 3.** Free body diagram and definitions.

Given that the rocket body is nearly axis-symmetric, the assumption of the angle of attack, $\alpha$, being zero leads to the assumption that lift, $L$, be also zero [16]. But other forces—the drag, $D$, the thrust, $T$, and the weight, $W$ (Fig. 3)—are variables that must be modeled with sufficient accuracy. Two forces, $T$ and $W$, are variable only during the thrusting phase, but $D$ is a variable during all three phases.

$$D = \frac{1}{2}\rho C_D A\, V_{\text{air}}^2 \quad (1)$$

Equation 1 shows a familiar expression for drag, $D$, where $A$ is the cross-sectional area (i.e. the frontal area) of the rocket body, $C_D$ is the drag coefficient, $V_{\text{air}}$ is the velocity of the rocket with respect to the air, and $\rho$ is the density of air. During the thrusting and the coasting phases, $C_D$ and $A$ are assumed constant. Thus, the variation in $D$ comes primarily from the variation in $V_{\text{air}}$.

We here note that the velocity with respect to the ground, $V_{\text{ground}}$, is the vector sum of $V_{\text{wind}}$ and $V_{\text{air}}$ [Fig. 3]. The pitch angle, $\theta$, is the angle between the ground and the rocket's body axis; the flight path angle, $\gamma$, is the angle between the trajectory of the rocket and the ground.

The density of air, $\rho$, can be calculated using the U.S. Standard Atmosphere [17]:

$$\rho = \rho_0 \left[1 - \frac{L_0 Y}{T_0}\right]^{1+gM/(RL_0)} \quad (2)$$

where $g$ is the standard free-fall acceleration, $L_0$ is the temperature lapse rate (assumed to be constant at $-0.0065$ K/m), $M$ is the molar mass of the air (assumed to be constant at 28.9645 g mol$^{-1}$), $R$ is the universal gas constant (8.3144598 J K$^{-1}$ mol$^{-1}$), $T_0$ is the temperature of the air on the ground (measured on the day of the launch), $Y$ is the altitude from sea level, $\rho$ is the density of the air, and $\rho_0$ is the density of the air at the ground.

*2.3. Equations of motion*

To predict the maximum altitude, the landing location, and the time of flight, the students analyzed the longitudinal dynamics of rockets while assuming that the motion of the rocket is confined to a plane perpendicular to the ground. They then numerically integrated equations of motion (Eqs. 3–6).

$$\frac{dV_{air}}{dt} = \frac{T}{m} - \frac{D}{m} - g\sin\theta \tag{3}$$

$$\frac{d\theta}{dt} = \frac{-g\cos\theta}{V_{air}} \tag{4}$$

$$\frac{dx}{dt} = V_{air}\cos\theta - V_{wind} \tag{5}$$

$$\frac{dy}{dt} = V_{air}\sin\theta \tag{6}$$

For numerical integration, students made use of MATLAB's built-in function called *ode45*, which is based on the classical fourth- and fifth-order Runge-Kutta methods [18, 19].

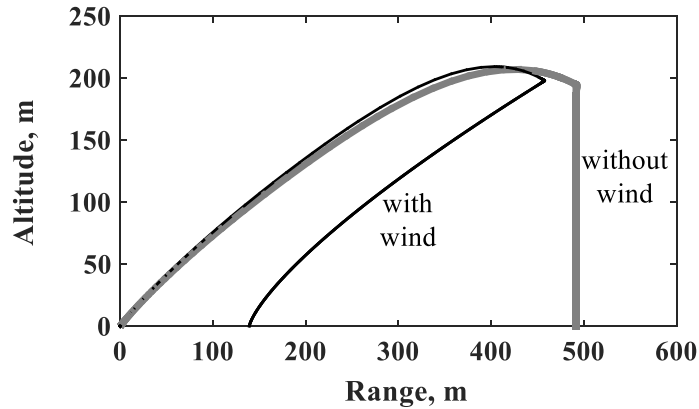

**Fig. 4.** Trajectory plots of two cases: with and without wind (1.5 m/s at 3 m height).

Figure 4 shows a simulated trajectory of the *Magician* rocket launched at 45° from the ground. The thick grey line corresponds to a case without wind, whereas thin black line corresponds to a case with wind. In the no-wind condition, the rocket would descend vertically once the parachute is deployed. When there is a wind, however, a horizontal drift could be significant. In this simulation, the students assumed a wind speed of 1.5 m/s measured at a height of 3 m. (As we will discuss later, the wind speed was assumed to change as a function of altitude; this is the reason that the descend trajectory for the drifting case would not result in a straight line.) Compared to the no-wind case, the landing location (for the assumed launch condition) may drift nearly 350 m to the downwind direction.

Page 6 of 22

## 3. Propellant analyses

When launched vertically, our *Magician* rocket would reach an altitude as high as 500 m (~1640 ft). The total flight duration of approximately 3 minutes would be enabled by the thrusting phase lasting mere 3 s. Adequately modeling these first 3 s is thus crucial in simulating the flight trajectory.

*3.1. The Estes E9-6 motor*

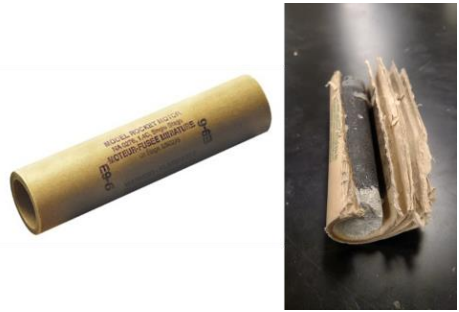

**Fig. 5.** Dissected E9-6 model rocket motor.

The Estes E9-6 rocket motor is shown in Fig. 5. Where the letter *E* is designated for the total impulse range of 20–40 Ns, *9* refers to an average thrust of 9 N, and *6* refers to a 6 s delay from engine burnout to the ejection charge ignition.

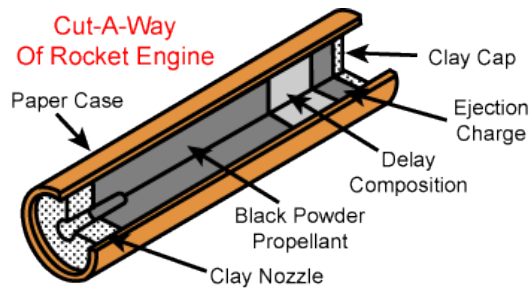

**Fig. 6.** Interior of a model rocket motor (credit: Apogee Components).

Figure 6 shows a schematic depicting an interior of a rocket motor. The majority of the internal volume is filled with propellant, presumed to be black powder. Once ignited, the burnt propellant would turn into gas, which would escape from the small hole of the nozzle cap made of clay. This hole is made small to maintain a high exhaust velocity. In case of the Estes E9-6 motor, this thrusting phase would last for 3 s. We see that the motor also contains delay composition. This charge would burn slowly to enable the coasting phase lasting for 4 s (±1 s) [20]. Finally, the ejection charge would ignite. The nearly instantaneous increase in the pressure inside of the rocket body would make the nose cone to slide out from the main body, causing the parachute to be deployed (Fig. 2).

*3.2. Infrared spectroscopy*

Our students proposed to estimate how much of the propellant's thermochemical energy was converted into kinetic energy. While the estimation of the kinetic energy of the exhaust gas could be achieved via an experiment (to be discussed later), the estimation of the thermochemical energy would not be possible without knowing the chemical composition of the propellant.

From the onset of the experiment, the students conjectured that the propellant in the Estes E9-6 motor was a black powder, i.e., some combination of potassium nitrate ($KNO_3$), charcoal (C), and sulfur (S). Over the course of history and depending on applications, different mixture ratios have been employed for black powder [21].

Students had to first confirm the presence of $KNO_3$, C, and S. For this, they proposed the use of infrared (IR) spectroscopy. The details of the IR spectroscopy performed by our students (Fig. 7) are discussed by Brewer et al. [2].

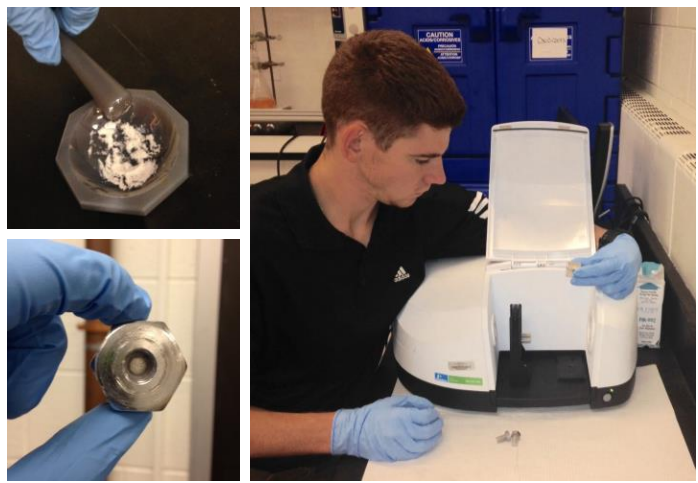

**Fig. 7.** Student performing infrared spectroscopy experiment.

The IR spectroscopy subjects a small sample of rocket propellant to a spectrum of the IR light, and records how much of that light is transmitted through the sample. The presence of a given compound would be confirmed by identifying wave numbers of the light being absorbed. Since $KNO_3$ is the primary ingredient in black powder, students would expect the sample to have lower transmittance at the wavenumbers at which $KNO_3$ absorbs light the most.

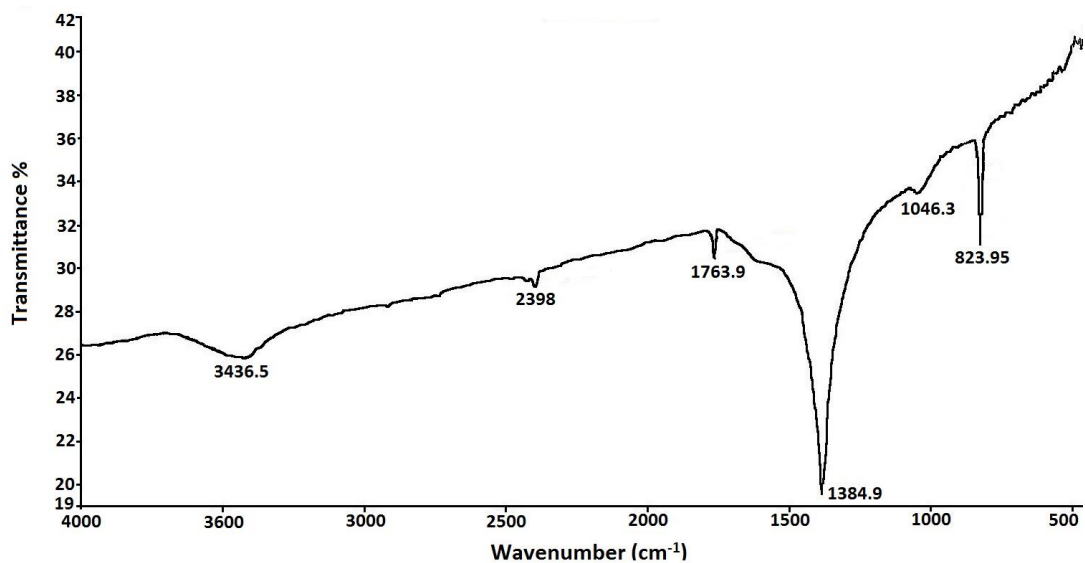

**Fig. 8.** The infrared spectrum of the model rocket propellant obtained by our students; the result matched to that expected for black powder.

The spectrum obtained by our students (Fig. 8) shows troughs in transmittance at wavenumbers 824, 1385, and 1764 cm$^{-1}$, which indicate the presence of nitrate—one of the main ingredients of black powder. Although not shown here, presence of sulfur and carbon could also be confirmed via additional IR spectroscopy of the burnt propellant [22].

*3.3. Extraction experiment*

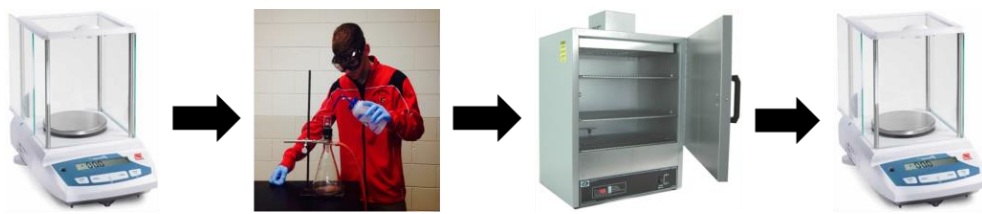

**Fig. 9.** Student performing one of the extractions.

After confirming that the propellant is made indeed of black powder, the students proceeded to determine the mixture ratio of potassium nitrate, sulfur, and carbon. To achieve this goal, the students consulted with the topic experts from the chemistry department, and conducted a series of extraction experiments.

Using the principle that different substances would dissolve in different chemicals, students could successively extract one chemical substance at a time (Fig. 9). By measuring the weight of the propellant sample before and after each extraction, the mass percentages of the extracted sample could be measured. The details of the procedure used by our students are presented by Brewer et al. [2]. Through this experiment, the mass percentages of potassium nitrate, sulfur, and carbon were measured to be 75.0%, 8.4%, and 16.2%, respectively. The result was consistent with that expected for a modern-day black powder.

The sum of the three components' mean concentrations was 99.89%, indicating that the experiments were carried out with sufficient accuracy. (The sum of the three substances was not expected to be perfectly 100% due to the small amount of moisture present in the propellant. Students conducted a separate experiment to confirm a presence of moisture, and found that the mass content of moisture was < 1% of the total mass of propellant [2]). A similar extraction experiment is discussed in the literature [23].

*3.4. Measurement of thrust profile and total impulse*

For their trajectory code, students needed an accurate measurement of how thrust from the rocket motor would change as a function of time. These thrust data were also necessary in the calculation of propellant efficiency.

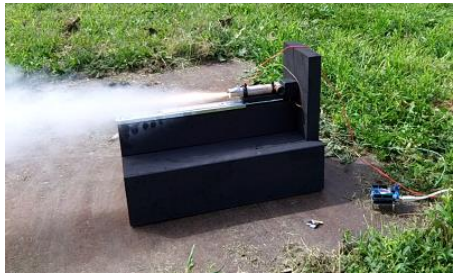 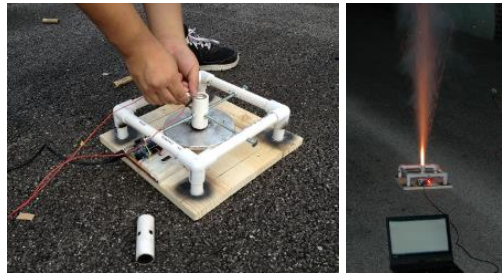

Credit: opendesignengine.net      Credit: Mike Szczys

**Fig. 10.** Examples of thrust measuring configurations.

The design of the thrust-measuring experiment would serve as a mini design project within the broader model rocket project. There are many potential design approaches for the rocket motor mounting apparatus, including configurations with motor mounted up, down, or sideways. The thrust could also be measured in a several different ways. Two examples are shown in Fig. 10.

The experimental design would be driven in part by the objectives and constraints set by the instructor and students. In our case, the entire experiment had to be designed using materials and equipment already available in the lab. In addition, the students proposed the apparatus for multiple purposes, such as measuring drag of rocket body and thrust of propellers used for radio-controlled airplanes.

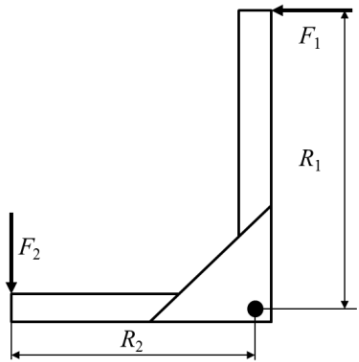

**Fig. 11.** Students' concept of rocket model thrust test apparatus.

Based on these considerations, the students elected to build an L-shaped apparatus (Fig. 11). The load was applied at one end of the arm and the measurement was taken at the other end of the arm. Because the lengths of the vertical and the horizontal arms were known, the thrust could be measured by balancing the moment (i.e., $F_1 R_1 = F_2 R_2$). If the expected load were too low or too high for the force sensor, the positions of the loading ($R_1$) or of the force sensor ($R_2$) could be adjusted.

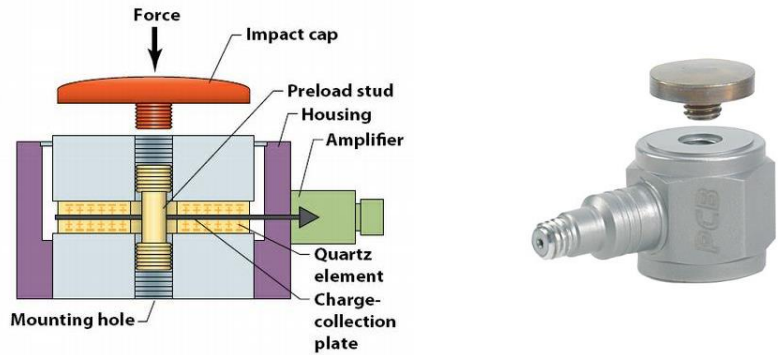

**Fig. 12.** Force sensor used by the students.

As for the force sensor, students used a load cell with a piezoelectric crystal (Fig. 12), which converts the applied force into an electrical signal.

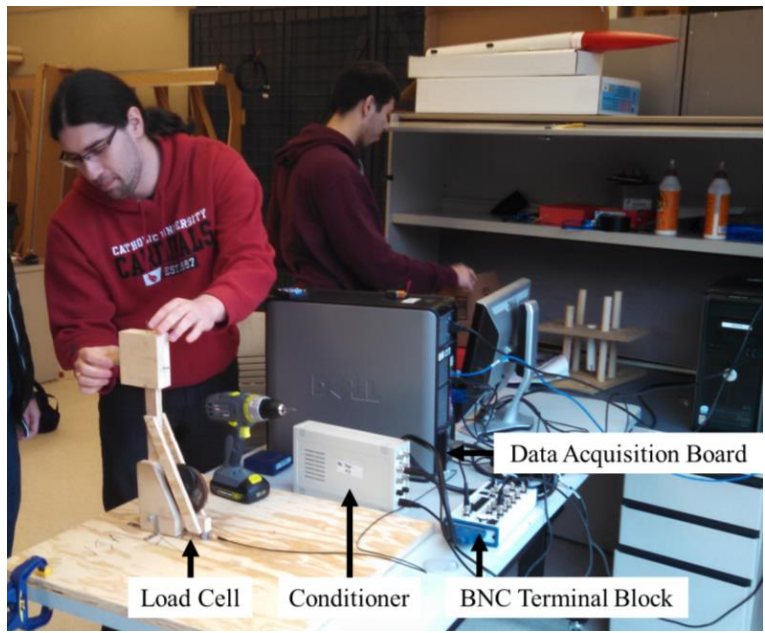

**Fig. 13.** Students setting up their thrust test apparatus.

Because the motor mounting apparatus and the load cell don't just record the data themselves, students had to configure the hardware and the software to enable the experiment. The white box in the middle of the worktable (shown in Fig. 13) is the signal conditioner, which amplifies, filters, and range matches the signal collected from the load cell. The signal conditioner is connected to a cable with the BNC (Bayonet Neil-Concelman) connectors, one end of which is connected to the BNC terminal block. The terminal block is used as an intermediate connector between the data acquisition board and the conditioner. The data acquisition board discretizes the analog signal into a digital signal. Students would write a LabView scripts to save this digital signal (collected a high sampling rate of 2 kHz) as a MATLAB data file.

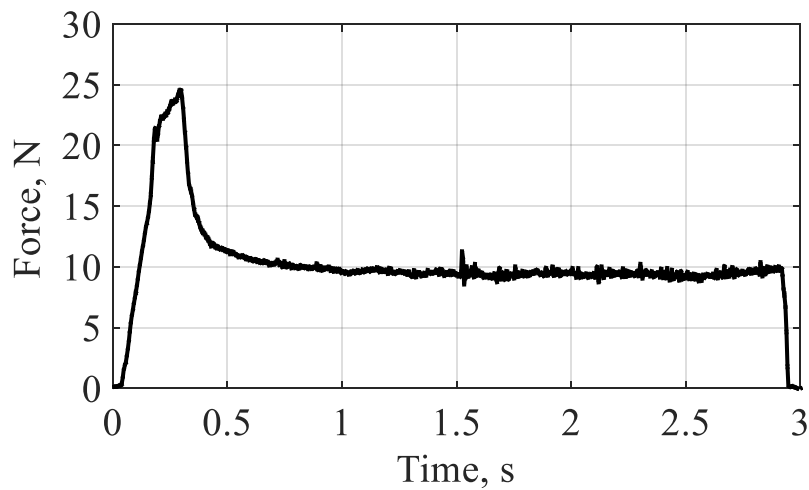

**Fig. 14.** Estes E9-6 thrust test by the students' exhibit expected results.

The National Association of Rocketry (NAR), which conducts independent testing of popular hobby rocket motors, reports that the Estes E9-6 motor used in our *Magician* rocket would have a peak thrust of 25 N (at around 0.2–0.3 s), followed by a nearly constant thrust at 10 N. The reported burn time is 3 s [20, 24]. The measurement obtained by our students (Fig. 14) matched this expectation. Their time-thrust curve was used as an input for their trajectory simulation and for calculation of propellant efficiency (using the fact that the area under the time-thrust curve represents the total impulse of the rocket motor).

Capturing the dynamic thrust reading was not as straightforward as measuring a static force (e.g., weight). Because there was no equipment already designed to measure the rocket motor's thrust, students had to think about a set of hardware and software that would together record the data in a usable format. A custom apparatus had to be designed and built. Setting up an experiment was a valuable educational exercise.

We presented the experiment carried out by our students merely as an example; it may not be an ideal approach if other hardware or software were available. For example, the piezoelectric crystal used in their load cell had an inherent problem of voltage discharge. This phenomenon would make a recorded force to appear to decay over time (even if the actual applied load remained constant). Because this voltage decay of the piezoelectric crystal follows a predictable exponential decay function, our students were able to calibrate the data by first measuring the decay constant and then multiplying the recorded data by an exponential function.

## 4. Estimation of drag coefficients

*4.1. Drag coefficients of the rocket bodies*

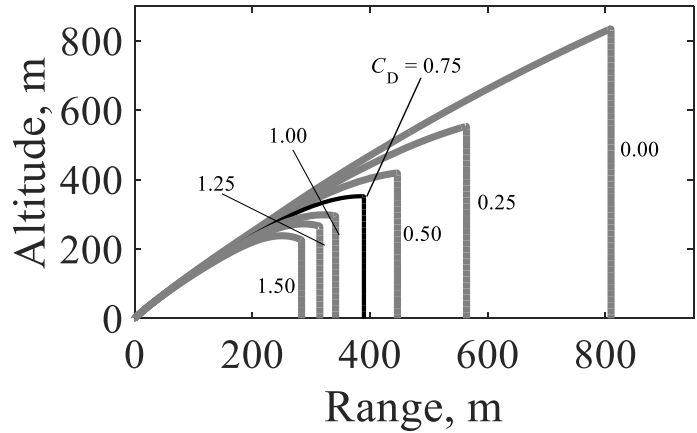

**Fig. 15.** Effects of $C_D$ values on the trajectories of rocket launched at 60° under no-wind condition.

The drag on rocket bodies has significant effect on the flight trajectories. Figure 15, for instance, shows how the $C_D$ values affect the trajectory of a rocket launched at 60°. In this plot, trajectory corresponding to $C_D = 0.75$ is shown in a thin black line. If the effects of atmosphere were ignored (by assuming $C_D = 0$), we the simulation would have incorrectly predicted that the rocket would reach an altitude twice as high and a range twice as far as the actual flight.

The students attempted to measure the $C_D$ value of the rocket body by constructing a small wind tunnel. Unfortunately, the results were found unreliable because the noise in the data was too high with respect to the magnitude of the value being measured. In the end, the students elected to use $C_D = 0.75$ (based on Ref. [25]) as a representative value.

*4.2. Drag coefficients of the parachutes*

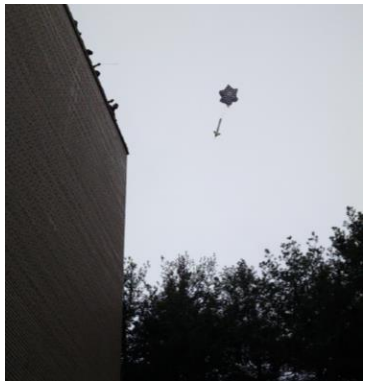

**Fig. 16.** The drag coefficient of a parachute-deployed configuration was determined via drop test.

Because the descending phase accounts for the majority of a rocket's time in flight, the drag coefficient of a parachute has significant effects on the landing location and the time of flight. To determine $C_D$, students conducted a drop test where a parachute (with rocket attached) was released from the rooftop of our engineering building (20 m tall). For every drop, multiple members clocked the time from release to impact. They used an average of multiple trials to estimate the descent time.

The descending motion of the rocket-parachute system is determined by the weight ($W$) pointing downwards and the drag ($D$) pointing upwards (ignoring the small drift due to wind, mostly in the horizontal direction). But because it was unclear how quickly the vehicle reaches its terminal velocity, the students used the differential equations of motion in order to identify the drag coefficient that matched with the experiment. The surface area of the hexagon-shaped parachute is $A = (\sqrt{3}/2)d^2$, where $d$ is the distance between two parallel sides. This area serves as reference geometry to allow $C_D$ to be compared against the literature. (We here note that, for reference areas, students used cross-sectional area for the rocket body and surface area for the parachute.)

**Table 1** Parachute Parameters

| Rocket name | Rocket burned-out mass, kg | Area of hexagonal parachute sheet, m$^2$ | $C_D A$ Measured in drop test, m$^2$ | Estimated $C_D$ |
|---|---|---|---|---|
| *Magician* | 0.121 | 0.181 | 0.231 | 1.28 |
| *Power Patrol* | 0.059 | 0.086 | 0.085 | 0.99 |

This drop test allowed the students to estimate the drag coefficient of the entire system (i.e. parachute, cords, and rocket fuselage). Their results are summarized in Table 1. Using the results from the drop test, they estimated the $C_D$'s of parachute-deployed configuration for the *Magician* and *Power Patrol* to be 1.3 and 1.0, respectively. As a point of comparison, Army's 35-foot T-10 personnel parachute system has $C_D$ of 0.8–0.9 [26]. (This $C_D$ value includes the aerodynamic effects of the payload and suspension lines.)

## 5. Estimation of wind velocity profile

Winds have a major impact on the flight trajectory. A linear approximation is sufficient to get the sense of this effect: with a typical flight duration of 2 minutes, an error of 1 m/s in the wind velocity would result in an error of 120 m in the landing location. An error in estimating the wind velocity would translate into a large error in predicted landing location. Our students faced challenges in both measuring and modeling the wind-velocity profile. As discussed later, this challenge remained an unsolved problem.

*5.1. Wind velocity profile model*

To estimate wind velocities for the entire range of altitudes flown by our rockets, students attempted to use a wind profile based on Ref. [27].

$$V_2 = V_1 \left(\frac{y_2}{y_1}\right)^a \quad \text{for} \quad V_2 > V_1, \quad y_2 > y_1 \tag{7}$$

where $V_1$ is the wind velocity at height $y_1$, $V_2$ is the wind velocity at height $y_2$, and $a$ is the Hellmann exponent. It must be noted that there is no scientific theory suggesting that the wind-velocity should vary according to this function.

As can be seen in Eq. 7, the model assumes that the wind velocity is zero at the ground level and increases with altitude. The rate of velocity increase, on the other hand, is high near the ground but reduces with altitude (where the effects of surface friction diminish). This wind model also assumes that the wind blows horizontally with no updrafts or downdrafts and that the wind direction is constant at all altitudes. The wind-velocity profile is assumed constant during the entire time of flight (as can be seen from the lack of a time variable in Eq. 7).

Different values of $a$ are used depending of the types of terrain. For instance, the value of $a$ may be 0.10 above an open water, but 0.34 above a rough terrain or urban area [27]. Unfortunately, a real launch site rarely fits into an idealized terrain category. Our launch site, for instance, was an open field with moderate hills, and was located near a residential neighborhood with plenty of treed areas. It was unclear which value of $a$ shall be used from the look-up table. To make the matter worse, the values of $a$ were observed to differ depending on the time of day and temperature, even for the same location [28].

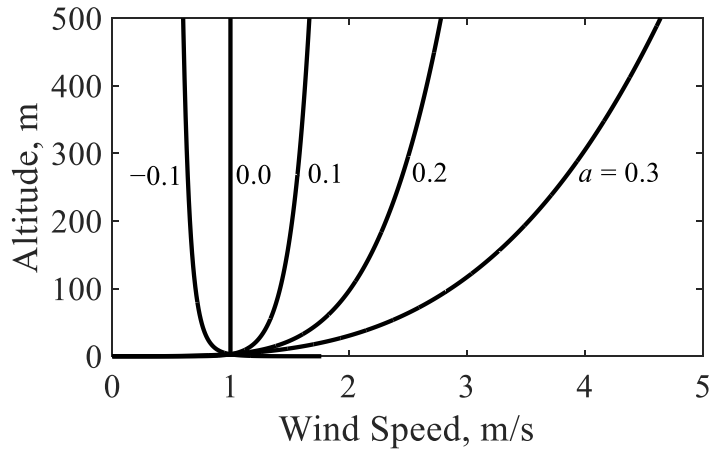

**Fig. 17.** A small difference in $a$ makes a large difference in predicted wind profile. All curves are based on a measured wind speed of 1 m/s taken 3 m above the ground.

The uncertainty in $a$ posed a major challenge, as this value controlled the shape of the wind profile curve and has a large effect in predicting velocity as a function of height. To illustrate this effect, students conducted a trade study, assuming the wind speed of 1 m/s measured at a height of 3 m (Fig. 17). We see in the figure that higher values of $a$ lead to slower increases in wind velocity with height (because rougher terrain causes more surface friction). We also see that that even a small difference in the value of $a$ (e.g., 0.2 vs. 0.3) would cause a large difference in the predicted wind speed (i.e., 2.8 vs. 4.6 m/s, respectively, an error of 39%) at the altitude of 500 m.

$$a = \frac{\ln(V_1/V_2)}{\ln(y_1/y_2)} \qquad (8)$$

Given the exponential nature of Eq. 7 being sensitive to the value of *a*, students had to measure the value of *a* at our specific launch site at the time of launch. Inverting Eq. 7 gives Eq. 8. The Hellmann exponent, *a*, could in theory be estimated if wind speeds at two different altitudes could be measured.

*5.2. Measurements of wind speeds*

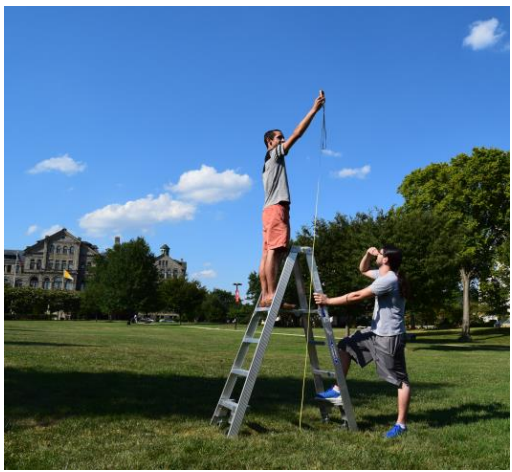

**Fig. 18.** Students demonstrate the measurement of wind using an anemometer held at a height of 3 m.

From the procedural point of view, measuring the wind speed near ground (i.e., $V_1$ in Eqns. 7 and 8) would be easy: students would simply face into the wind and record the reading from an anemometer held at a certain height from the ground. (The photo in Fig. 18 was taken for the purpose of demonstration, as the actual rocket launch was carried out in an open field.) What is not easy was how to use this measured speed for trajectory simulation. Rocket's flight is not an instantaneous event. Rather, it takes place over 2–3 minutes, during which the wind speeds (and directions) fluctuate, often considerably. Thus, improving the accuracy of an instantaneous wind measurement would not improve the accuracy of simulation.

The difficulty of modeling the effects of wind was compounded by the difficulty of estimating the wind speed at a higher altitude (i.e., $V_2$). To minimize the error in estimating the value of *a*, it would be prudent to set the $y_2$ value to be higher than the maximum altitude reached by the rocket, so that the values would be estimated via interpolation, rather than extrapolation. But how shall students measure the wind speed at the height of > 500 m? Measurement using a hand-held anemometer would require them to climb up on a building taller than the Empire State Building. Unfortunately, there was no such skyscraper near our launch site, because the site was chosen specifically as to be away from buildings.

One practical option for the wind speed data was *Winds Aloft* (i.e., Winds and Temperature Aloft Forecast), issued by the National Oceanic and Atmospheric Administration (NOAA) and the National Weather Service. Because *Winds Aloft* fortuitously included the weather data at at the location of our launch site (i.e., Westminster, Maryland), the reported wind speed at 3000 feet (914 m) seemed like a good candidate for the $y_2$ value in Eq. 8 (which was used to estimate *a*). Unfortunately, *Winds Aloft* also had serious drawbacks: the reported wind speeds are forecasts, not real-time data, and the values are updated only four times a day.

The anemometer reading at 3 m and the *Winds-Aloft* data at 3000 feet could not be used to estimate the value of Hellmann exponent, *a*. When the students checked the reliability of this approach by repeating the procedure every day for a period of one week, the estimated *a* varied from as low as −0.12 to as high as 0.23, even though all estimates were done at the same time of the day. In the end, the students adopted a compromise, and simply assumed an *a* value of 0.20, corresponding to "agricultural land with trees" [27].

## 6. Measurement of maximum altitude

On the day of the launch, students measured the landing location, the time of flight, and the maximum altitude. These results were compared against their predictions. They used a tape measure to measure the landing location and a stopwatch to measure the time of flight. Measuring the maximum altitude, however, required some preparations.

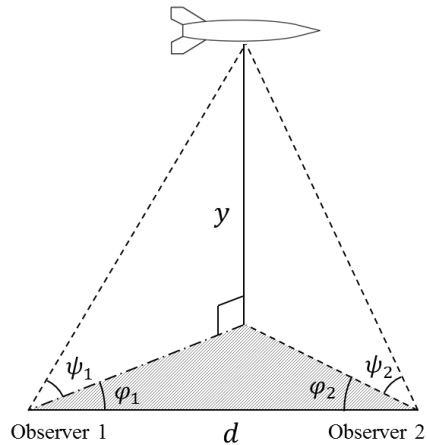

**Fig. 19.** The altitude of rocket was calculated from the angles measured using theodolites.

The maximum altitude of the rockets can be calculated if the angles between two observers and the rocket were known (Fig. 19). Two observers each would measure the vertical angle between the ground and the rocket, and the horizontal angle between the rocket and the other observer. The subscripts 1 and 2 in Fig. 19 denote the observers 1 and 2, respectively.

Based on the geometry, the altitude, *y*, could be estimated using one of the following relationships [29].

$$y = \frac{d \tan\psi_1 \tan\psi_2}{\cos\varphi_1 \tan\psi_2 + \cos\varphi_2 \tan\psi_1} \tag{9}$$

$$y = \frac{d \tan\psi_2 \sin\varphi_1}{\sin(\varphi_1 + \varphi_2)} \tag{10}$$

$$y = \frac{d \tan\psi_1 \sin\varphi_2}{\sin(\varphi_1 + \varphi_2)} \tag{11}$$

where *d* is the distance between the observers, *y* is the altitude measured from the ground, *φ* is the horizontal angle between the rocket and the other observer, and *ψ* is the vertical angle of elevation between the ground and the rocket. To reduce the error in estimates, the student team used an average of the *y* values obtained from all three equations.

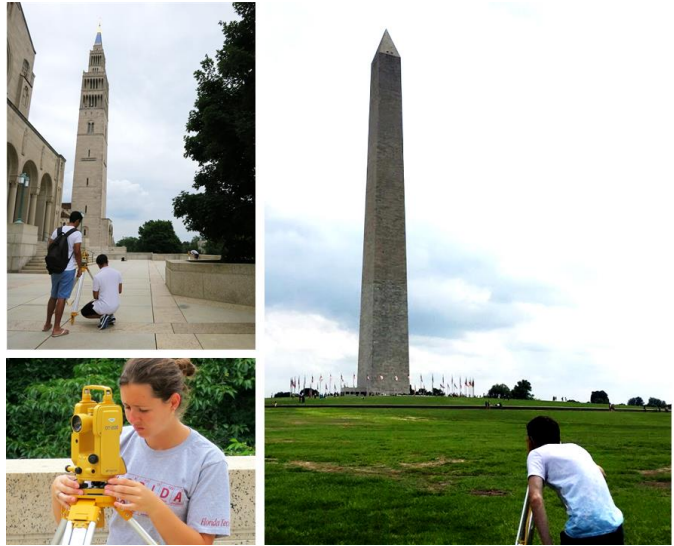

**Fig. 20.** Students verifying the altitude estimation techniques by measuring objects of known heights.

Our students learned how to use theodolites, a surveying tool used by civil engineers. Most engineering schools with civil engineering programs would have this equipment (Fig. 20). Prior to the launch date, the students validated their method by measuring several buildings of known heights, including the bell tower on campus and the Washington Monument in Washington D.C. (Fig. 20).

The use of Eq. 9–11 was found to be a reliable way to measure height. For instance, students measured the height of the Washington Monument to be 168.989 m. Assuming that 169.046 m (+/− 1.0 mm) measured in the National Geodetic Survey [29], which was conducted by the National Oceanic and Atmospheric Administration, to be the correct height, their measurement error was only 0.057 m or 0.034%.

While theory and equipment for the height measurement method were accurate when applied to building structures, applying the method to a fast-moving object posed additional challenge. The rocket could disappear into the cloud. The sun could obstruct the view of the observers. And even if the observer could track the rocket—the trail of smoke made this task easier—identifying when, exactly, it has reached its maximum altitude was often impossible to know. Thus, when comparing the accuracy of trajectory calculation against measurements taken during the flight test, the measurement of the altitude itself was assumed to have uncertainties. We could confirm the inaccuracy in the altitude measurement from the inconsistencies between values calculated from Eq. 9–11. If the measurements were perfect, these three equations would have shown an identical value.

The telescope on the theodolite could be rotated horizontally and vertically. But these telescopes could not be used at a near vertical angle, so the equipment must be placed far enough from the location where the maximum altitude would be reached. [There is another reason to avoid a near-vertical viewing angle: when *ψ* are close to 90°, a small error in *ψ* would result in a large error in the predicted altitude due to the tan($\psi$) term in the equation.] On our launch day the theodolites were positioned more than 250 m away from the launch pad.

## 7. Additional data recording

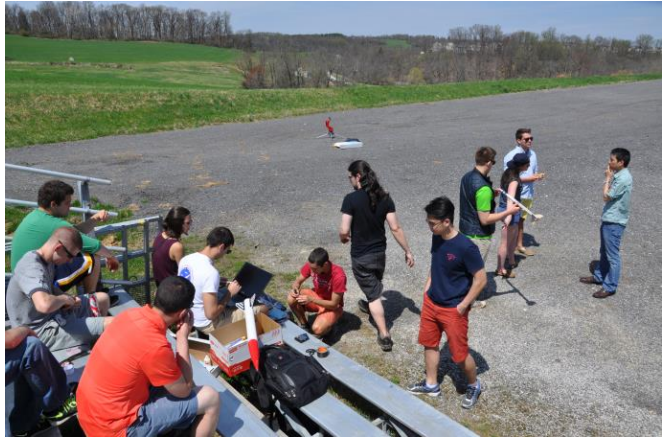

**Fig. 21.** Students re-run their simulation codes using the on-site wind and temperature measurements as input data.

We shall expect continuing enhancements in model rocket projects, as electronic devices continue to become smaller, cheaper, and more widely available. One of the rockets flown by our students, for instance, had a barometry-based altimeter, which would have provided an additional and independent measurement of the maximum altitude reached by the rocket. (Unfortunately, we could not include the result in this paper, as this particular vehicle was lost in the nearby forest.)

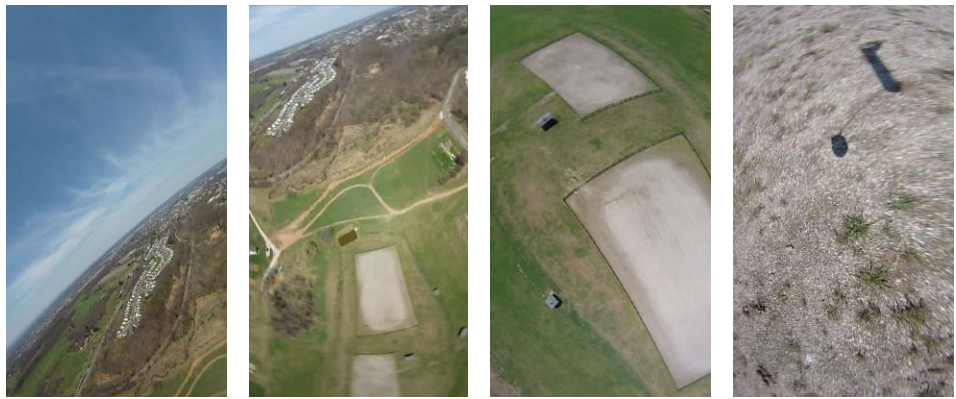

**Fig. 22.** On board video camera chronicles rocket's flight from launch to landing.

Portable cameras have reached a practical size and cost that they are now suitable for model rockets flown as class projects. Fig. 22 displays images from video footage taken from a camera affixed to the nose of a rocket flown by our students.

# 8. Conclusions

Analyses, simulations, and experiments associated with model rockets present themselves as exciting and practical educational experiences. The multi-disciplinary nature of the project offers an opportunity to integrate knowledge and skills students learned in a variety of undergraduate courses. As the camera and other digital devices becoming smaller and cheaper, model rockets would continue to evolve as compelling projects that could be adopted in most engineering schools.

## Acknowledgments

This paper is based on a class project for ME554 Aerospace Design (Instructor: Masataka Okutsu) at The Catholic University of America (CUA), Washington, D.C. The first author served as project manager of the student team. We thank Ronaldo Chavez Reis, Ronaldo Limberger Tomiozzo, John Brewer, and Spencer Seufert for their technical contributions and Margaret S. Morris for her editorial suggestions. We are grateful to Sen Nieh, chair of the Department of Mechanical Engineering at CUA, for his support and encouragement.